# Fast response and highly sensitive flexible humidity sensor based on nanocomposite film of $MoS_2$ and graphene oxide

Gengwu Ge,[1,2] Ningfeng Ke,[1,2,4] Hongliang Ma,[1,2,3] Jie Ding,[2*] Wendong Zhang,[5,6*] and Xuge Fan[1,2,3*]、

[1]Advanced Research Institute for Multidisciplinary Science, Beijing Institute of Technology, Beijing 100081, China.

[2]Center for Interdisciplinary Science of Optical Quantum and NEMS Integration, Beijing Institute of Technology, 100081 Beijing, China.

[3]School of Integrated Circuits and Electronics, Beijing Institute of Technology, Beijing 100081, China.

[4]Yangtze Delta Region Academy of Beijing Institute of Technology, Jiaxing 314003, China.

[5]State Key Laboratory of Dynamic Measurement Technology, North University of China, Taiyuan 030051, China.

[6]National Key Laboratory for Electronic Measurement Technology, School of Instrument and Electronics, North University of China, Taiyuan 030051, China.

*Email: xgfan@bit.edu.cn, jie.ding@bit.edu.cn, wdzhang@nuc.edu.cn

**Abstract**

Graphene oxide (GO)-based humidity sensors are attracting widespread attention due to their high responsivity and low cost. However, GO-based humidity sensors generally suffer from slow response and recovery as well as poor stability, etc. Here, we reported a flexible resistive humidity sensor based on a $MoS_2$/GO composite film that was fabricated by mixing different volumes of $MoS_2$ and GO dispersions with adjustable volume ratios. The $MoS_2$/GO composite film has been used as a sensing layer on screen-printed interdigital electrodes. The results show that the best device performance was achieved at a dispersion volume of 0.05 mL with the $MoS_2$/GO volume ratio of 5:1, featuring high responsivity (~98%), fast response/recovery time (1.3/12.1 s), excellent stability and low cost. Further, the humidity sensor exhibits good linearity over a wide humidity range (33% RH-98% RH) at room temperature (25°C) and can be fabricated easily and feasibly. The application of the humidity sensors we prepared in human respiration detection and human fingertip proximity detection has been demonstrated. These findings indicate the great potential of the composite of $MoS_2$/GO in developing the next generation of high-performance humidity sensors.

## Introduction

Humidity has always been an essential indicator for industrial production, medical and health care, environmental monitoring, and household use.[1–3] Along with the rapid development of modern industry, network communication, Internet of Things (IoT) technology, medical monitoring and health management, the demand for humidity detection is also growing.[4,5] There are several different humidity sensing mechanisms,[6] such as resistive humidity sensors,[7–9] capacitive humidity sensors,[10–12] field-effect transistor humidity sensors,[13] etc. Among them, resistive humidity sensors[14–17] have been the primary target of research on various humidity sensors due to easy preparation and intuitive electrical output form. Human respiration produces a large amount of water vapor, which can be captured for respiratory monitoring,[18,19] and cardiovascular and pulmonary diseases to provide a basis for judgment.

Molybdenum disulfide ($MoS_2$)[20–25] has high electrocatalytic activity, large specific surface area, and a wealth of surface-responsive sites, which has excellent potential to be used in the field of sensors[26,27] (Fig. 1a). However, devices based on pure $MoS_2$ without a specific process has poor repeatability after work for a few cycles. Graphene oxide (GO) that is the oxidation product of graphene[28–30] has excellent hydrophilic properties due to the abundance of oxygen-containing functional groups (including hydroxyl, epoxy, and carboxylic acid groups) on its surface and can provide large surface-responsive sites due to its large surface volume and unique two-dimensional structure (Fig. 1b). Therefore, the GO-based humidity sensors are highly sensitive. However, the pure GO has no apparent response to relative humidity change.

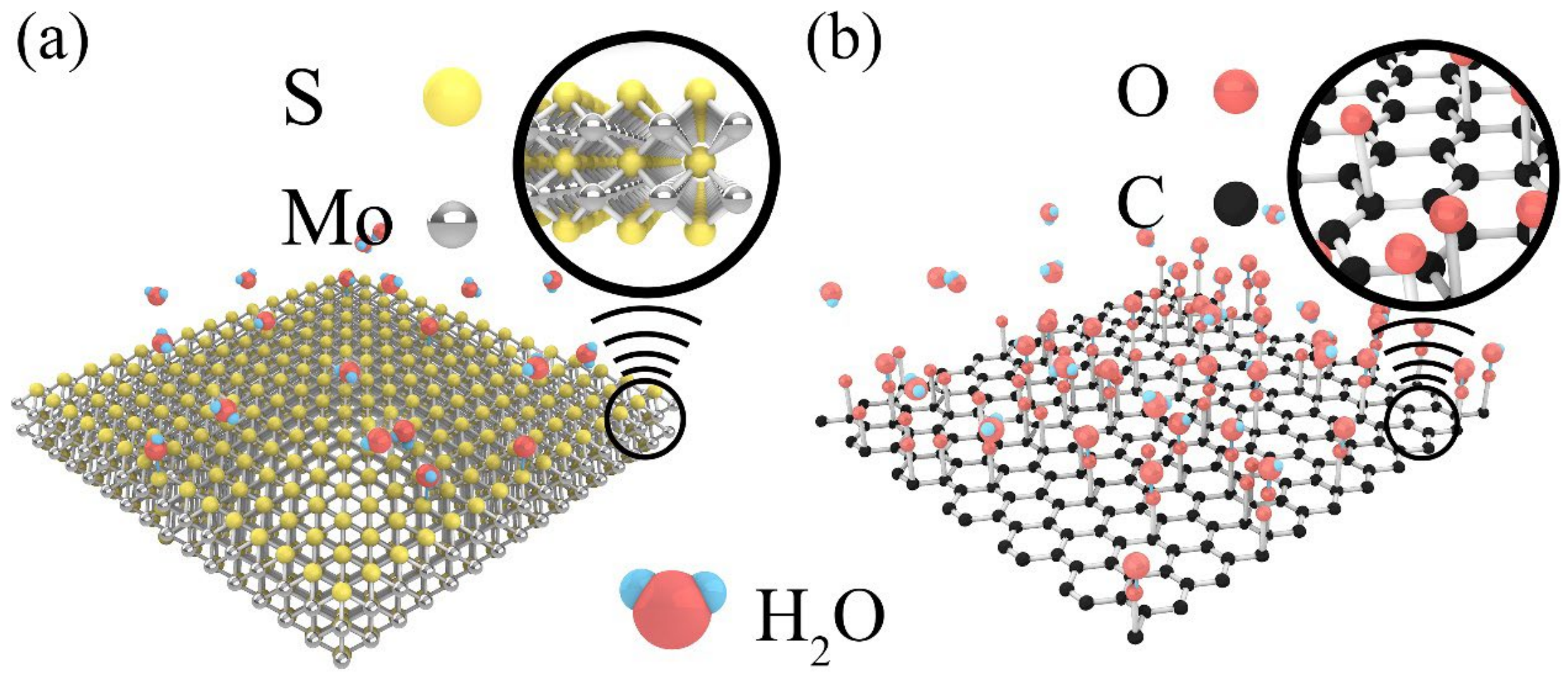

**Fig. 1** Schematics of $MoS_2$ and GO. (a) Structural schematic of MoS2. (b) Structural schematic of GO.

Humidity sensors based on pure $MoS_2$ or pure GO are limited in either high responsivity or fast response and recovery time. Pure $MoS_2$ film has no stable repeatability in the long term for stable daily use. In addition, there are few reports of humidity sensors based on the pristine molybdenum disulfide films. Nevertheless, treated $MoS_2$ has excellent humidity sensing performance and related works have been reported in recent years. Burman et al. reported a resistive humidity sensor based on Pt-decorated $MoS_2$ nanoflakes in 2018,[31] showing a high responsivity of 97.5% in the range of 25%RH-85%RH but a long response time (larger than 90 s) and recovery time (larger than 150 s). Zhao et al. reported a capacitance humidity sensor based on a $SnO_2$-modified $MoS_2$ in 2016,[32] which shows an excellent response time (17 s) and recovery time (6 s) but low responsivity (31.37%). Jin et al. reported an inkjet-printed $MoS_2$/PVP hybrid humidity sensor in 2014,[33] featuring a responsivity of 45.02%, response/recovery time of 90 and 100 s, respectively.

GO was also reported to be used as a sensitive film for humidity sensing by using the optimized process. Alrammouz et al. reported a highly porous and flexible capacitive humidity sensor based on self-assembled GO in 2019,[34] showing a responsivity of 52.15% in the relative humidity range of 30% RH-90% RH but a pretty long response time (200 s) and recovery time (100 s). From now on, fewer studies have been reported on humidity sensors based on $MoS_2$ and GO composite films. Burman et al. reported a humidity sensor based on $MoS_2$/GO composite films in 2016,[35] exhibiting a responsivity of about 94.12% in the relative humidity range of 35%RH to 85%RH and a response/recovery time of 43 and 37 s, respectively. However, it seems difficult for the humidity sensor based on $MoS_2$/GO composite films to meet the demand for high responsivity and fast response/recovery time simultaneously. [35]

In this work, we focus on a new flexible resistive humidity sensor based on composites of $MoS_2$/GO with high responsivity, fast response/recovery time, and excellent stability. The fabrication process is based on the layer-by-layer drop-coating method, and thereby, the humidity sensor can be realized quickly and feasibly. The typical advantages of the fabricated humidity sensor include high responsivity (98%), fast response time and recovery time (1.3/12.1s), and superior stability. The prepared humidity sensor featured good linearity over a wide humidity range (33% RH-98% RH) and low cost. The fabricated humidity sensors were used for human respiratory monitoring and human fingertip proximity detection, showing excellent performance.

## Experimental section

### Preparation of $MoS_2$/GO dispersions

The humidity-sensitive materials mentioned in this work were prepared, as shown in Fig. 2a-c. The raw materials were $MoS_2$ dispersion with thin-layer $MoS_2$ (1 mg/mL, aqueous solvent) and GO dispersion with thin-layer GO (2 mg/mL, aqueous solvent) purchased from XFNANO (Nanjing, China). The used volume ratios of $MoS_2$ to GO are 1:1, 3:1, 5:1, 7:1, and 10:1. Then the dispersion of $MoS_2$ and GO were ultrasonicated at 30°C for 60 minutes, respectively. Consequently, they were mixed well[36]. Finally, the brownish-black aqueous dispersion of $MoS_2$/GO was obtained after natural cooling at room temperature.

### Preparation of flexible PET substrate with interdigital electrode (IDT)

As shown in Fig. 2d, the surface of 100 μm thick PET film was treated with plasma to enhance the adhesive strength between the PET substrate and the conductive silver paste. Second, a screen-printing mask of the interdigital electrode was customized and used to prepare the interdigital electrode of conductive silver paste on the PET film by the drop coating method. The scraper was used to ensure the uniform thickness and continuous plane of conductive silver paste during the drop coating process. After removing the screen-printing mask, the conductive silver paste on the PET substrate was heated at 120°C for 20 minutes to sinter. Finally, the interdigital electrode on the PET substrate was obtained. The conductive silver paste solution with a silver particle content of 68% was used, and the sheet resistance of the conductive silver paste was not larger than 0.2 Ω.

**Preparation of flexible $MoS_2$/GO based humidity sensor**

The humidity sensor based on $MoS_2$/GO composite is prepared by layer-by-layer drop coating method (Fig. 2 e-g). Specifically, the prepared flexible PET substrate with an interdigital electrode was fixed on a stainless-steel disk and put on a 95°C homogenized hot plate. The 50 μL brownish-black aqueous dispersion of $MoS_2$/GO was drop-coated on the prepared flexible PET substrate. After the $MoS_2$/GO film was dried, another 50 μL dispersion of $MoS_2$/GO was drop-coated on the previous dry $MoS_2$/GO film. This process repeated back and forth until the $MoS_2$/GO dispersion was used up.

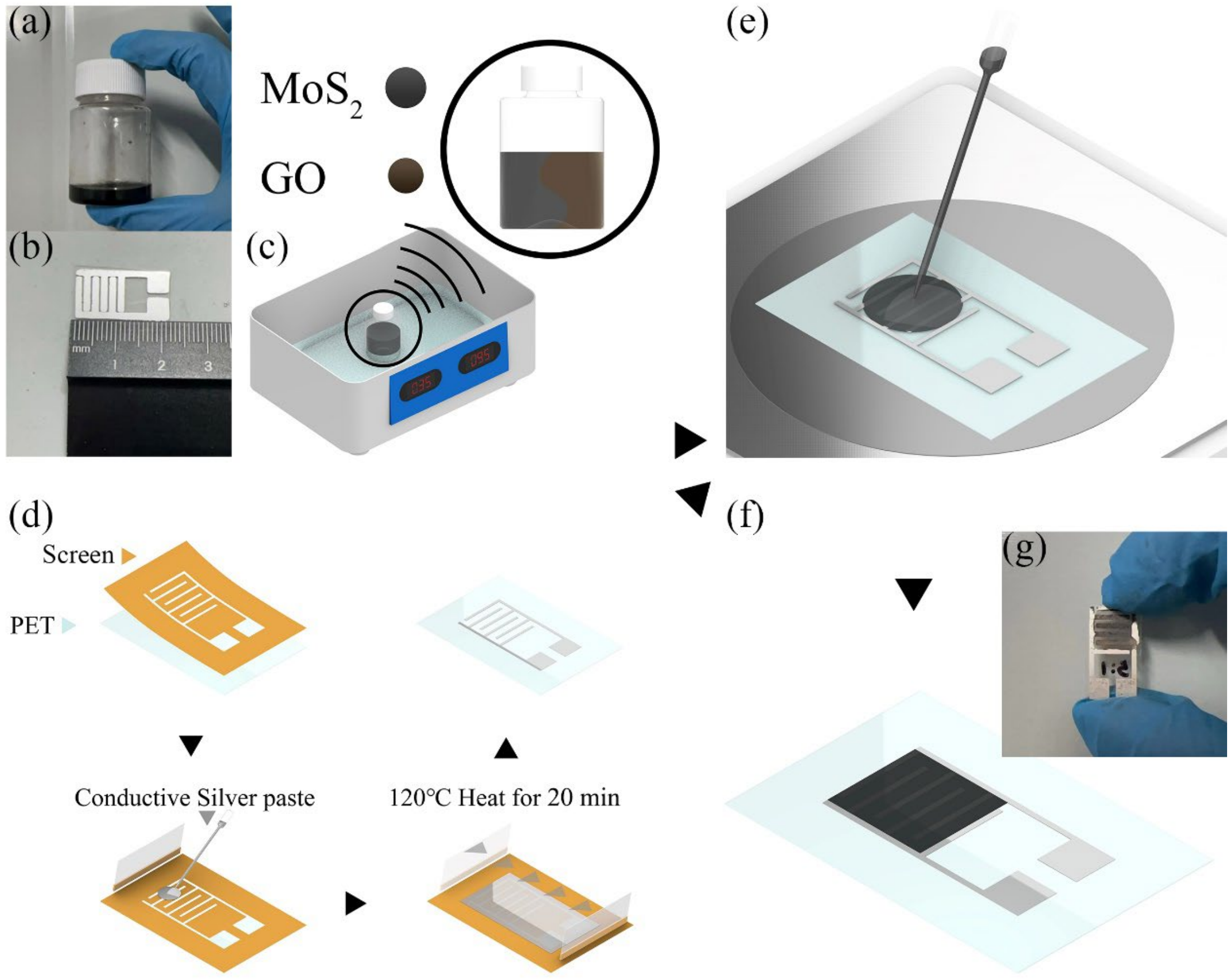

**Fig. 2** Fabrication process of $MoS_2$/GO composite-based humidity sensor. (a, b) Mixed $MoS_2$/GO dispersion and printed interdigital electrodes. (c) $MoS_2$/GO mixed dispersion ultrasonicated for 60 min at 30°C. (d) Preparation of finger electrodes. (e) Drop the ultrasonicated dispersion of $MoS_2$/GO on top of the prepared silver finger electrode and heat it at 95°C. (f) Schematic of humidity sensor based on $MoS_2$/GO composite. (g) A photo of the fabricated $MoS_2$/GO composite-based humidity sensor with a volume ratio 5:1.

**Characterization of $MoS_2$, GO and $MoS_2$/GO composite films**

As shown in Fig. 3, the surface morphology of the prepared pure $MoS_2$ films (Fig. 3 a-c) and pure GO films (Fig. 3 d-f), as well as the $MoS_2$/GO composite films (Fig. 3 g-h), were observed by optical microscope (Olympus BX53M) and scanning electron microscopy (SEM) (Zeiss Supra 55). The prepared pure $MoS_2$ samples exhibited uniform distribution of grainy structure (Fig. 3 a-c). By contrast, the prepared pure GO samples featured the distribution of film structure due to the agglomeration effect and thereby had strong moisture absorption ability due to the increased surface area (Fig. 3 d-f). As shown in Fig. 3 g-h, the $MoS_2$ and GO can be distinguished from the prepared $MoS_2$/GO composite films, in which dark agglomerations with the large areas were GO due to the agglomeration effect while scattered granular structures were $MoS_2$. More SEM and optical microscope images of the prepared $MoS_2$/GO composite films can be seen in Fig. S1 and Fig. S2 in Supporting Information. The preliminary electrical characterization of the prepared $MoS_2$/GO composite films on flexible PET substrate was performed by a probe station connected to a Source Meter (Keithley 2450, Tektronix). The initial resistance values of the prepared humidity sensor samples can be seen in Table S1 and Table S2 in the Supporting Information.

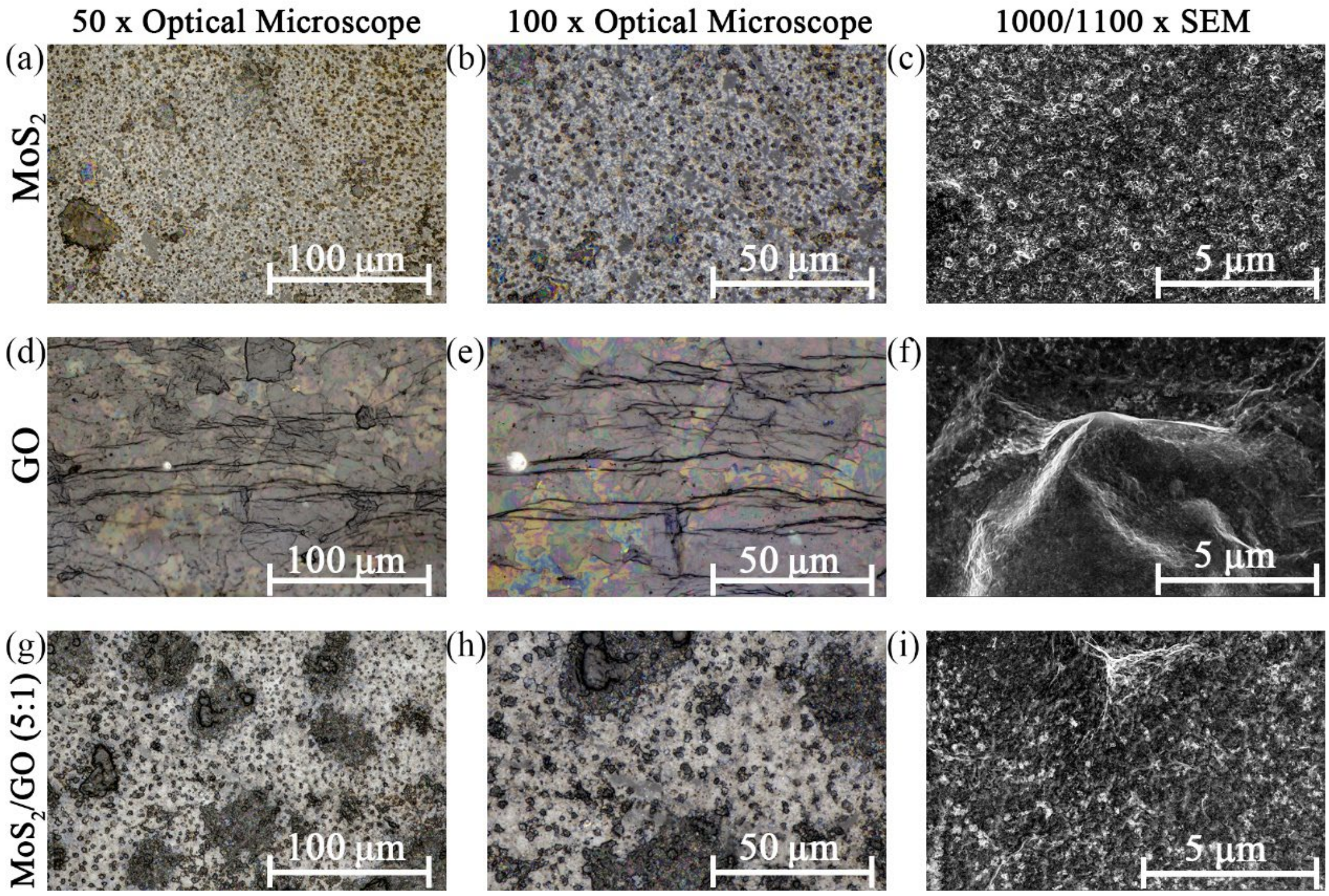

**Fig. 3** Optical and SEM characterizations. (a, b) Optical images and (c) SEM image of prepared pure $MoS_2$ films on the flexible PET substrate. (d, e) Optical images and (f) SEM image of prepared pure GO films on the flexible PET substrate. (g, h) Optical images and (i) SEM images of the prepared $MoS_2$/GO composite films on the flexible PET substrate with a volume ratio of 5:1.

**Humidity measurements**

Based on previous reports[37,38], the humidity environments with the relative humidity of 33%RH, 43%RH, 59%RH, 75%RH, 85%RH, and 98%RH were obtained by saturated salt solution of $MgCl_2$, $K_2CO_3$, NaBr, NaCl, KCl, and $CuSO_4$, respectively. All measurements are performed at the room temperature (25°C). The ambient humidity is 11% RH. The measurements of response and recovery time were performed in a 98%RH microenvironment. The resistance changes of the prepared flexible humidity sensors after exposure to different humidity environments were measured by a digital multifunctional multimeter (Keithley DAQ6510, Tektronix).

**Performance**

In this work, the responsivity is defined by Eq. (1) below:

$$Responsivity = \frac{|R_1 - R_0|}{R_0} \times 100\% \qquad (1)$$

Where $R_0$ is the resistance of the sensor under the condition of ambient humidity (11%RH), and $R_1$ is the resistance of the sensor after exposure to the specific humidity condition.

The responsivity normalized by relative humidity change can be obtained by responsivity divided by the relative humidity change. The response time was defined as the time that is taken when the resistance change of the sensor is increased from 0 to 90%, and the recovery time was the opposite. The prepared humidity sensors were firstly placed in an ambient humidity environment of 11% RH and then put in a humidity environment of 98% RH for 2 minutes to measure the responsivity and response time. After that, the prepared humidity sensors were put back into the ambient humidity environment (11% RH) for 2 minutes to measure the recovery time.

**Results and discussion**

**Humidity sensing performance**

To determine the volume of $MoS_2$/GO dispersion for the best performance of flexible humidity sensors, 50 μL, 100 μL, 300 μL, 500 μL, 700 μL, and 1 mL of $MoS_2$/GO dispersions with the $MoS_2$/GO volume ratio of 1:1 were prepared in this work and were drop coated on the PET substrate for humidity measurements. Thus, by determining the optimal volume of $MoS_2$/GO dispersion, the best performance of the humidity sensor was obtained. Then, the usage of the volume of $MoS_2$/GO dispersion will be fixed, in which the $MoS_2$/GO volume ratio of 1:1, 3:1, 5:1, 7:1, and 10:1 will be chosen for the preparation of $MoS_2$/GO composites and humidity measurements. As comparisons, the dispersion consisting only of $MoS_2$ and GO was also separately used to prepare $MoS_2$ (Fig. 3 a-c) and GO films (Fig. 3 d-f) on flexible PET substrates, respectively, and consequently for humidity measurements.

As the relative humidity changed from 11% RH to 98% RH, the resistance changes of the prepared humidity sensors decreased with the increase of the volume of $MoS_2$/GO dispersion (Fig. 4 (a)). The relative resistance changes of the prepared humidity sensors generally decreased with the increase of the volume of $MoS_2$/GO dispersion except for some random fluctuations (Fig. 4 b). As shown in Fig. 4 b, the highest responsivity (42.4%) was obtained from the humidity sensor sample that was prepared by using the volume of 0.05 mL $MoS_2$/GO dispersion with a fixed $MoS_2$/GO volume ratio of 1:1.

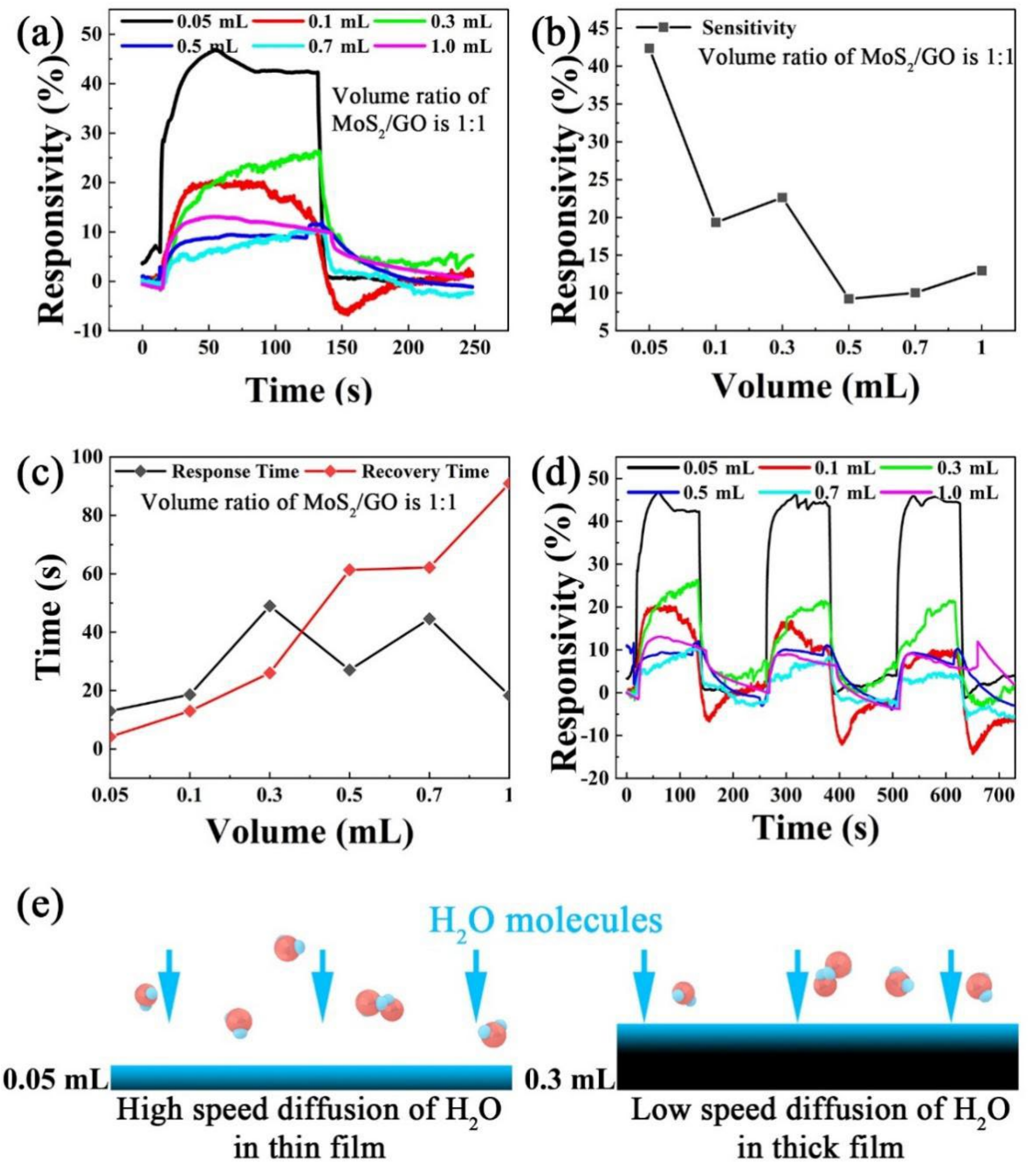

**Fig. 4** Humidity measurements of fabricated flexible humidity sensors based on different volumes of $MoS_2$/GO with volume ratio of 1:1. (a) Response and recovery curves of samples that were prepared by using different volumes of $MoS_2$/GO with volume ratio of 1:1. (b) Responsivity of samples in (a). (c) Response time and recovery time of samples in (a). (d) Repeatability of samples in (a) for three cycles of humidity measurements. (e) Schematic of different diffusion speeds of water molecules in $MoS_2$/GO composite films with different thicknesses that were prepared by using 0.05 mL $MoS_2$/GO dispersion (thin films) and 0.3 mL $MoS_2$/GO dispersion (thick films) in high humidity environment.

Further, the fastest response time (13.1 s) and recovery time (4.2 s) were also obtained from the same humidity sensor sample (Fig. 4 c), indicating that the responsivity decreases while the recovery time increases as the volume of $MoS_2$/GO dispersion increases. The recovery time appears to have a significant increase as the volume of $MoS_2$/GO dispersion increases, especially when the dispersion volume is at 0.5 mL. The thicker $MoS_2$/GO composite films' recovery time is longer than the thinner $MoS_2$/GO composite films.

For the humidity sensors with thinner $MoS_2$/GO composite films, the water molecules can easily and rapidly interact with the surface of thinner $MoS_2$/GO composite films as the samples are moved from a low humidity environment (11% RH) to a high humidity environment (98% RH). For the humidity sensors with thicker $MoS_2$/GO composite films, it takes a longer time for the water molecules to diffuse from the surface of the $MoS_2$/GO composite film to the bottom of the composite film. Therefore, the response times for the thicker $MoS_2$/GO composite films are generally longer than those of thinner $MoS_2$/GO composite films. However, the response time was not always increased with the increase of the volume of the $MoS_2$/GO dispersion but fluctuated within a specific range of values (Fig. 4c). This might be probably ascribed to the gradually increased surface areas of the lateral walls of the $MoS_2$/GO composite film as the volume of the $MoS_2$/GO dispersion was increased. That is to say, although the $MoS_2$/GO composite film became thicker and thicker with the increased volume of the $MoS_2$/GO dispersion, the surface areas of the lateral walls of the $MoS_2$/GO composite film that have the ability to adsorb the water molecules were also continuously increased, which would contribute to the fast response to the relative humidity to some extent. Another possible reason is that as the volume of the $MoS_2$/GO dispersion was increased, the pristine resistance of the $MoS_2$/GO composite decreased, which probably resulted in the reduced amount of water molecules that are required for the $MoS_2$/GO composite to reach the minimum resistance of the $MoS_2$/GO composite and thereby contributed to the short response time to the humidity.

Three cycles of humidity measurements of humidity sensors prepared based on different volumes of $MoS_2$/GO dispersions were performed at relative humidity ranging from 11% RH to 98% RH (Fig. 4 d). The results show that the as-prepared humidity sensors exhibit excellent repeatability for most of the volumes of $MoS_2$/GO dispersion that were used. The humidity sensors also feature excellent stability after being put in different humidity environments for seven days. As shown in Fig. 4 b and c, the humidity sensors show the best performance as 0.05 mL $MoS_2$/GO dispersion was used. Therefore, the 0.05 mL $MoS_2$/GO dispersion will be used for consequent experiments including preparation of $MoS_2$/GO composite films and humidity measurements.

The 0.05 mL dispersions with different $MoS_2$/GO volume ratios (1:1, 3:1, 5:1, 7:1, 10:1) were used for the preparation of the $MoS_2$/GO composite films, and consequent humidity measurements of prepared humidity sensors were performed (Fig. 5 a). As shown in Fig. 5 b and c, the best responsivity (about 98%) and response time (1.3s) were obtained from the humidity sensors made of 0.05 mL dispersion with $MoS_2$/GO volume ratio of 5:1. The corresponding responsivity normalized by the relative humidity change was 1.51%/RH%. Compared with $MoS_2$/GO volume ratio of 5:1, the humidity sensors based on other $MoS_2$/GO volume ratios show decreased responsivities and increased response times, whatever higher or lower than 5:1. In addition, humidity sensors based on 0.05 mL dispersion with $MoS_2$/GO volume ratio of 5:1 shows a recovery time of 12.1 s, which is higher than most of humidity sensors based on other $MoS_2$/GO volume ratios (Fig. 5 c).

When the $MoS_2$/GO volume ratio is 1:1, the responsivity of the sensor is low. This is because there are not enough $MoS_2$ molecules to produce sufficient resistance changes in humidity sensors after exposure to the humidity environment. The response time gradually improves as the percentage of $MoS_2$ increases until the $MoS_2$/GO volume ratio reaches 5:1. In addition, as the volume ratio of $MoS_2$/GO is 5:1, the prepared humidity sensor showed the best responsivity and excellent repeatability and stability besides the shortest response time (Fig. 5d). For the sharp decrease in responsivity as the ratio of $MoS_2$ to GO increases beyond 5:1, we believe that as the $MoS_2$/GO volume ratios are larger than 5:1, the content of GO in the dispersion of $MoS_2$/GO is so low that the water molecules captured by the GO molecules are not enough and couldn't cover all the $MoS_2$ molecules between GO molecules, which results in bad performance in terms of responsivity as well as response time and recovery time. Because water molecules adsorbed on the surface of $MoS_2$ molecules can be directly desorbed as the sample was moved from the high humidity environment to the low humidity environment, the desorption performance of the humidity sensors seems not be significantly affected by the volume ratio of $MoS_2$/GO.

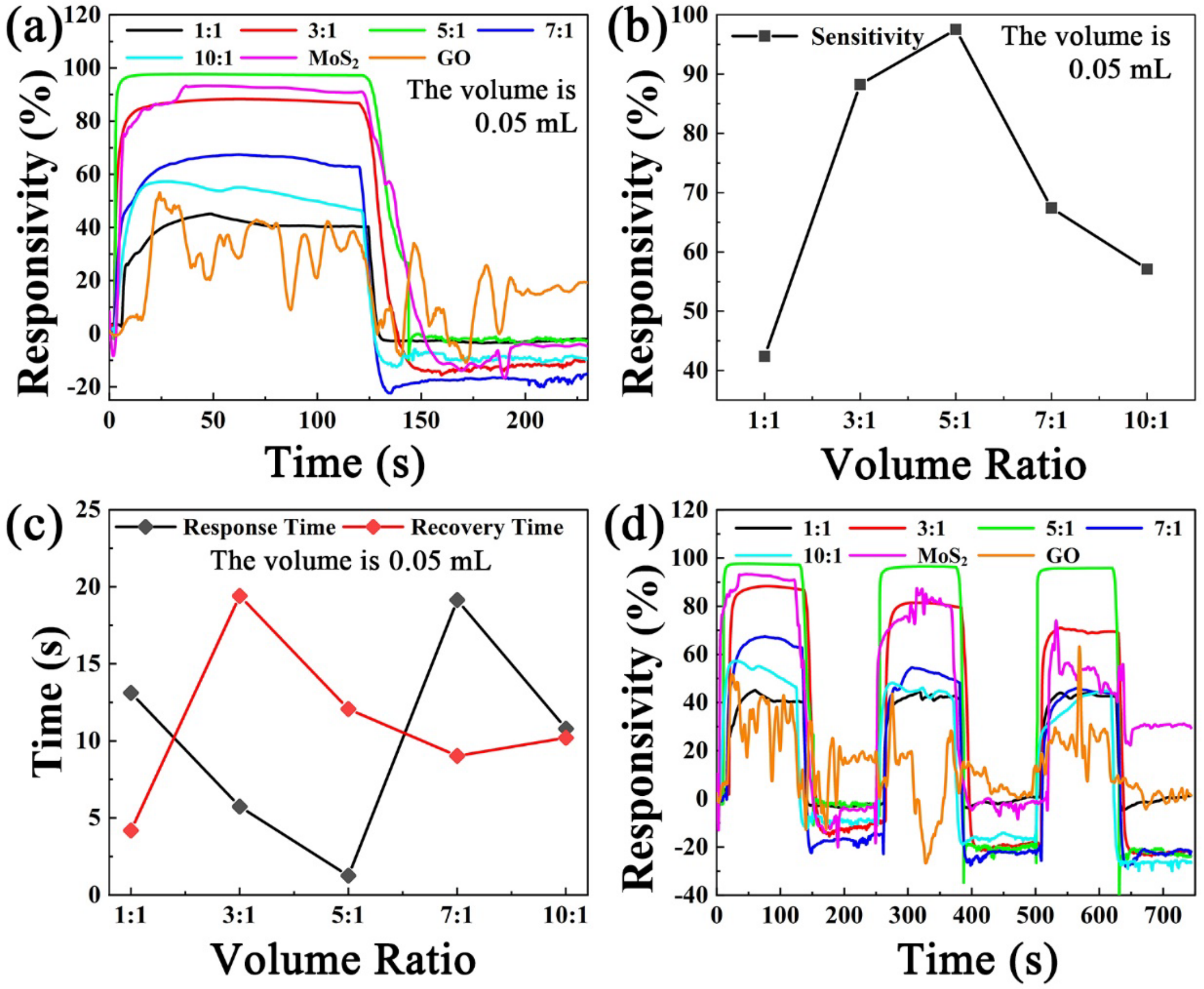

**Fig. 5** Humidity measurements of fabricated flexible humidity sensors based on 0.05 mL pure $MoS_2$ dispersion, 0.05 mL pure GO dispersion, and 0.05 mL $MoS_2$/GO dispersion with different volume ratios. (a) Response and recovery curves of samples were prepared using 0.05 mL pure $MoS_2$ dispersion, 0.05 mL pure GO dispersion and 0.05 mL $MoS_2$/GO dispersion with different volume ratios of 1:1, 3:1, 5:1, 7:1, and 10:1. (b) Responsivity of samples in (a). (c) Response time and recovery time of samples in (a). (d) Repeatability of samples that were prepared by using 0.05 mL pure $MoS_2$ dispersion, 0.05 mL pure GO dispersion, and 0.05 mL $MoS_2$/GO dispersion with different volume ratios of 1:1, 3:1, 5:1, 7:1, and 10:1 for three cycles of humidity measurements.

For comparisons, humidity sensors based on pure $MoS_2$ and pure GO were also prepared by using 0.05 mL $MoS_2$ dispersion and 0.05 mL GO dispersion, respectively. It should be noted that the responsivity of the sample based on pure $MoS_2$ was lower than the sample based on $MoS_2$/GO volume ratio of 5:1 but higher than other samples based on all other $MoS_2$/GO volume ratios (Fig. 5 a). This might because the pristine resistance of the sample based on pure $MoS_2$ is lower than the sample based on the $MoS_2$/GO volume ratio of 5:1 but higher than other samples based on all other $MoS_2$/GO volume ratios (Table S2). Another possible reason we speculate is that $MoS_2$ particles were uniformly distributed around the GO films for the sample based on $MoS_2$/GO volume ratio of 5:1, resulting in the formation of the adjacent heterojunctions one by one, which would contribute to the best responsivity.

Fig. 5 d shows three cycles of measurement results of prepared humidity sensors based on 0.05 mL pure $MoS_2$ dispersion, 0.05 mL pure GO dispersion, and 0.05 mL dispersion with different $MoS_2$/GO volume ratios, respectively. As shown in Fig. 5 d, a humidity sensor made of 0.05 mL pure $MoS_2$ dispersion shows an excellent responsivity of over 90%, a response time of 1.2 s, and a recovery time of 18 s (Fig. 5 a). However, after two cycles of humidity measurements, the responsivity significantly decreased (Fig. 5 d), indicating that the humidity sensors based on pure $MoS_2$ film were unreliable as the humidity environment is changed at a fast speed, with a bad repeatability and stability. On the other hand, the humidity sensor made of 0.05 mL pure GO dispersion shows quite low responsivities with large fluctuations, whether one cycle of humidity measurements or three cycles of humidity measurements (Fig. 5 a and d).

It can be seen from Fig. 5 a and d that the performance of humidity sensors based on pure $MoS_2$ films is unreliable in the humidity environment with fast changes. This is because $MoS_2$ does not have good hygroscopic and desorptive properties. Likewise, the electrical output signal of humidity sensors based on pure GO films is quite unreliable. This is because the prepared pure GO samples were typically presented by the folded film structures due to agglomeration effect of GO, which probably resulted in bad contact with the Ag interdigital electrodes and thereby resulted in fluctuation of the electrical output signal. While, the GO has large number of oxygen-containing groups that can provide a large number of water adsorption sites. By adding a certain proportion of GO in the $MoS_2$ dispersion to form $MoS_2$/GO composite films, the relatively good hygroscopic and desorption properties of GO are able to help $MoS_2$ films respond better to humidity, which ultimately improves the overall performance of humidity sensors.

It can be concluded that the humidity sensor sample using 0.05 mL $MoS_2$/GO dispersion with a volume ratio of 5:1 to prepare composite film on the PET substrate shows the best humidity sensing performances in terms of responsivity, response time and recovery time, stability, and repeatability. It should be noted that if less than 0.05 ml $MoS_2$/GO dispersion was used, it was difficult to obtain the high-quality, uniform and continuous $MoS_2$/GO composite film on the flexible PET substrate with the interdigital electrodes. Table 1 compares our work with previous literature about humidity sensors based on different types of humidity-sensing materials. After comparison, the response time (1.3 s) of our humidity sensors is significantly better than others while the recovery time of our humidity sensors is also better than most of previously reported humidity sensors.

The absolute resistance of the prepared humidity sensors changed with the time during three cycles of humidity measurements can be seen in Fig. S3 a and b in Supporting Information. And the measurements of the response time and recovery time in detail can be seen in Fig. S3 c and d in Supporting Information. To further demonstrate the excellent stability and repeatability of the prepared humidity sensor using 0.05 mL $MoS_2$/GO dispersion with a volume ratio of 5:1, we performed the humidity experiments with the time intervals of one day, two days, three days, five days, ten days and half month, respectively and found that the changes of the responsivity of the sample that was measured on different days can be ignored.

**Table 1.** Performance of the prepared sensor compared with previous study

| Type | Sensing material | Preparation Methods | Measurement range (%RH) | Responsivity normalized by humidity (%/%RH)* | Res./ Rec. time (s) | Linearity | Ref. |
|---|---|---|---|---|---|---|---|
| Resistance | fMWCNTS/HAGO | Drop coating | 35-95 | 1.23 | 8/45 | Nonlinear | 39 |
| TFBG | GO/MWCNTs | Dip coating | 30-90 | 0.377 | 4/- | 35-65% Linear 65-90% Nonlinear | 40 |
| Voltage | GO | Drop casting | 33-98 | / | 0.28/0.3 | 33-70% Linear 70-98% Linear | 41 |
| Resistance | GO/PCF | Hydrothermal | 11-97 | / | 19/28 | Nonlinear | 42 |
| Resistance | $MoS_2$-flakes | Aerosol printing | 10-95 | 1.18 | 8/22 | Nonlinear | 43 |
| Resistance | rGO/$MoS_2$ | Drop casting | 5-85 | 0.29 | 6.3/30.8 | Linear | 44 |
| Resistance | Pt decorated $MoS_2$ | Dip coating | 25-85 | 1.63 | 91/154 | Nonlinear | 31 |
| Resistance | $MoS_2$/GO | Drop casting | 35-85 | 1.88 | 43/37 | Low linearity | 35 |
| Resistance | $SnO_2$/rGO | Dip coating | 11-97 | 0.52 | ~90/~100 | Linear | 38 |
| Resistance | $MoS_2$/PVP | Inkjet printing | 11-94 | 0.96 | 5/2 | Nonlinear | [33] |
| Capacitance | $SnO_2$-modified $MoS_2$ | Drop coating | 0-90 | 0.35 | 17/6 | Nonlinear | [32] |
| Capacitance | GO | Dip coating | 30-90 | 0.87 | ~200/~100 | Nonlinear | [34] |
| Capacitance | ITO/Alumina | Screen printing | 5-95 | 0.96 | 47.2/49.5 | Nonlinear | [45] |
| Resistance | $MoS_2$/GO | Drop coating | 33-98 | 1.51 | 1.3/12.1 | Linear | This work |

* The responsivity was obtained by $\Delta R/R_0$ for the resistive humidity sensors. It was obtained by $\Delta C/C_0$ for the capacitive humidity sensors, where the $R_0$ stands for the initial resistance of the device, and $C_0$ stands for the initial capacitance of the device. The responsivity normalized by humidity was obtained by $\Delta R/R_0$ divided by relative humidity change or $\Delta C/C_0$ divided by relative humidity change.

The humidity sensors prepared by using 0.05 mL $MoS_2$/GO dispersion with the volume ratio of 5:1 were used to perform the humidity step measurements. To be specific, the experiment was carried out by putting a humidity sensor sample sequentially into different bottles with different saturated salt solutions that produce different values of relative humidity (Fig. 6 a). The humidity sensor sample was put in each bottle with different saturated salt solutions for one minute and moved gradually from high humidity environment (98%RH) to low humidity environment (33%RH) and then back to high humidity environment (98%RH).

As shown in Fig. 6 b, the $MoS_2$/GO based humidity sensors show good stability during the process of humidity step experiment, which indicates that the sensor has excellent humidity absorption and desorption performance. Fig. 6 c shows the resistance of the $MoS_2$/GO based humidity sensor increased from about 80 kΩ to about 700 kΩ as the relative humidity is decreased from 98%RH to 33%RH that is a desorption process of water molecules. Its resistance decreased from about 700 kΩ to about 80 kΩ with insignificant hysteresis as the relative humidity increased from 33% to 98%RH, which is an absorption process of water molecules. At low RH levels, where water molecules are scarce in the air, fewer water molecules are adsorbed on the surface of the sensor, resulting in a less pronounced change in response. At high RH levels, the sensor surface is saturated with water molecules, and the response change tends to level off. At medium RH levels, however, the response change of the sensor is most significant because the number of adsorbed water molecules increases substantially at this time, which significantly affects the electrical properties of the sensor (Table 1). Nevertheless, the linearity of GO-based humidity sensors can be effectively improved by rationally adjusting the ratio of GO to $MoS_2$ during the process of preparation of $MoS_2$/GO composite. As shown in Fig. 6 d, data points of resistance versus relative humidity during absorption process of water molecules were fitted and the results of curve fitting show that the resistance of the humidity sensor has a highly linear relationship ($R^2$ = 0.95502) with relative humidity, which can be shown by Eq. (2) below:

$$y = (968575) + (-9455.47)x \quad (2)$$

where y is the resistance of the humidity sensor, and x is the relative humidity.

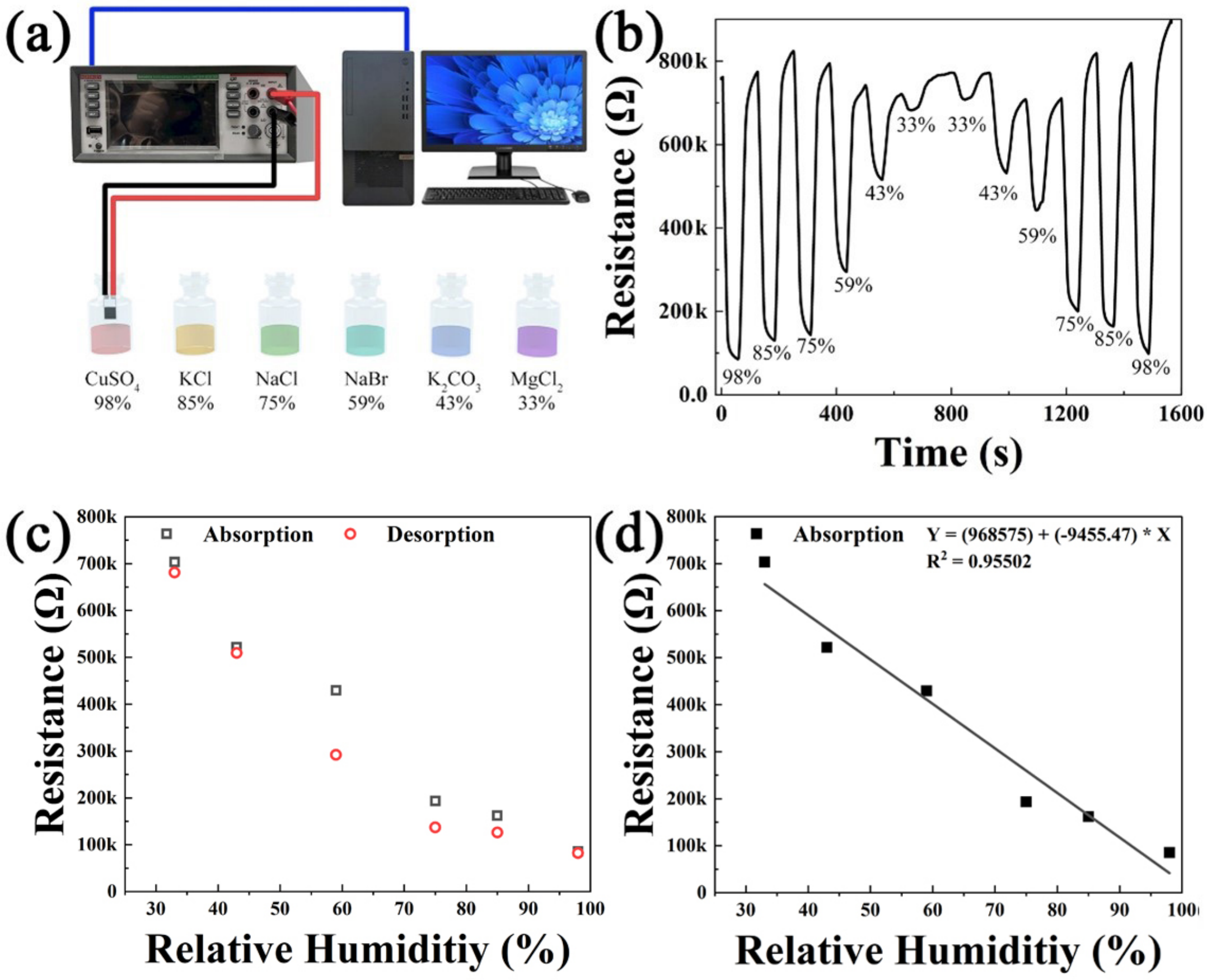

**Fig. 6** Humidity step experiment. (a) Schematic of the process of humidity step experiment. (b) The curve of the resistance of the prepared $MoS_2$/GO based humidity sensor changes with the time. (c) Comparison of the resistance of the prepared $MoS_2$/GO based humidity sensor, in which the humidity was decreased from 98%RH to 33%RH and then back to 98%RH. (d) Curve fitting of data points of resistance versus relative humidity during the absorption process of water molecules increased the humidity from 33%RH to 98%RH.

**The discussion of the humidity sensing mechanism**

The main humidity sensing mechanism is generally believed to be ion exchange between the sensitive material and adsorbed water molecules, which was proposed by Grotthuss[46,47]. This humidity sensing mechanism based on ion exchange can be expressed by Eq. (3) below, which is helpful to the rapid transfer of ions in a continuous water layer.

$H_2O+H_3O^+ \rightarrow H_3O^+ + H_2O$[31,48] (3)

Based on the optical microscope and SEM images (Fig. 3) of the $MoS_2$/GO composites, it can be inferred that the GO film covers the surface of $MoS_2$. Doping GO into $MoS_2$ to form $MoS_2$/GO composite can improve the humidity sensing performance of $MoS_2$ such as its responsivity and stability, by using the high moisture absorption capacity of GO.

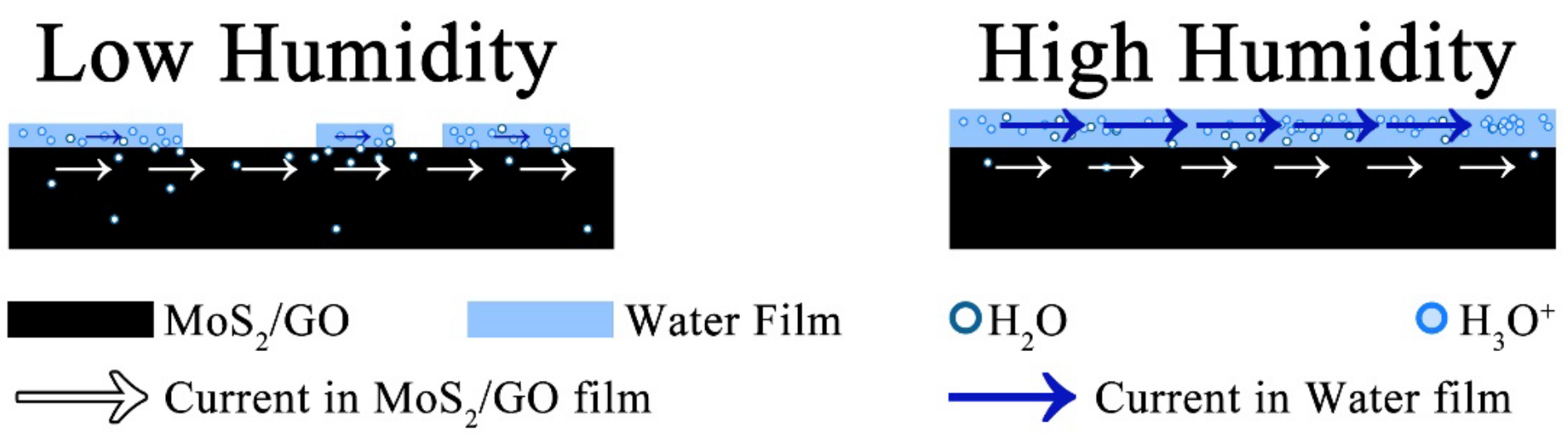

**Fig. 6** Schematic of the current on the surface of the prepared humidity sensors based on $MoS_2$/GO composite film in low and high humidity environment, respectively.

In a low relatively humidity environment, the water molecules might only adsorb on the partial top layer of the $MoS_2$/GO composite film and are difficult to form a continuous water molecule layer on the surface of $MoS_2$/GO composite films, which hinders the smooth transfer of $H_2O$ or $H_3O^+$ in the non-continuous water molecule layer (Fig. 7). Thus, $MoS_2$/GO composite films shows a higher impedance in a low relatively humidity environment. In a high relative humidity environment, the water molecules easily form a continuous water layer on the surface of the $MoS_2$/GO composite film, which accelerates the transfer of $H_2O$ or $H_3O^+$ in the continuous water layer (Fig. 7). Thus, $MoS_2$/GO composite films show the reduced impedance in a high relatively humidity environment.

**Application of prepared flexible humidity sensors**

For the human respiration monitoring, the prepared humidity sensors were put in a position 3 cm distance away from the human nose and mouth, which is the typical distance for a human to wear the mask. As shown in Fig. 8 a and b, the human respiration monitoring was carried out in different respiration rates, including 3s inhalation and 3 s exhalation (slow mode of breathing), 2 s inhalation and 2 s exhalation (middle mode of breathing), as well as 1s inhalation and 1s exhalation (fast mode of breathing). The resistance change of the humidity sensors were recorded to show the response curves in respect to the breathing behaviors. The results show that the resistance change of the prepared humidity sensor is consistent with the breathing rate, showing quite a good response to the breathing. Further, the prepared humidity sensors have higher resistance responsivity and better stability for human respiration monitoring in slow mode of breathing, compared to middle and fast mode of breathing.

When the 1 s inhalation and 1s exhalation experiments were carried out, the resistance of humidity sensors showed a visible drift.

Two types of experiments were performed for the detection of human fingertip proximity. First, boxes with an area of 1cm×1cm and different heights (0.1 cm, 0.3 cm, 0.5 cm, 0.7 cm and 1 cm) were prepared and put on top of the $MoS_2$/GO composite films for consequent humidity measurements in a confined space. One measurement cycle includes that the fingertip was put on top of boxes with different heights for 30 seconds and then the fingertip was away from the top of the boxes for 30 seconds. As shown in Fig. 8 c, the humidity sensor has a significant and stable resistance response to the fingertip after several measurement cycles and the resistance change of the humidity sensor increases with the decrease of distance of the finger away from the surface of the $MoS_2$/GO composite films. Second, the fingertip was directly put on top of the $MoS_2$/GO composite films with the distances of 0.1 cm, 0.3 cm, 0.5 cm, 0.7 cm, and 1 cm for humidity measurements in an open space. Likewise, the fingertip was put on top of $MoS_2$/GO composite films with different heights for 30 s and then the fingertip was away from $MoS_2$/GO composite films for 30 s. As shown in Fig. 8 d, the humidity sensor has a significant resistance response to the fingertip, and the resistance change of the humidity sensor obviously depends on the distance between the finger and the $MoS_2$/GO composite films. That is, the smaller the distance, the larger the resistance change. The responsivity of the humidity sensors versus the distance between the finger and $MoS_2$/GO composite films in confined space and open space were shown in Fig. 8 e and f, respectively. The responsivities of the humidity sensors decrease with the increase of the distance between the finger and the $MoS_2$/GO composite films, for whatever confined space or open space. To further study the relationship of the responsivity with the distance between the finger and the $MoS_2$/GO composite films, curve fittings were performed, which shows that the responsivity has quadratic function with the distance. Especially when located in open space, the sensor shows a linear relationship ($R^2 = 0.915296$) with the distance of the fingertip, which can be expressed by Eq. (4) below:

$$y = 83.1419 + (-90.3918) * x \quad (4)$$

Where y is responsivity, and x is the distance of the fingertip.

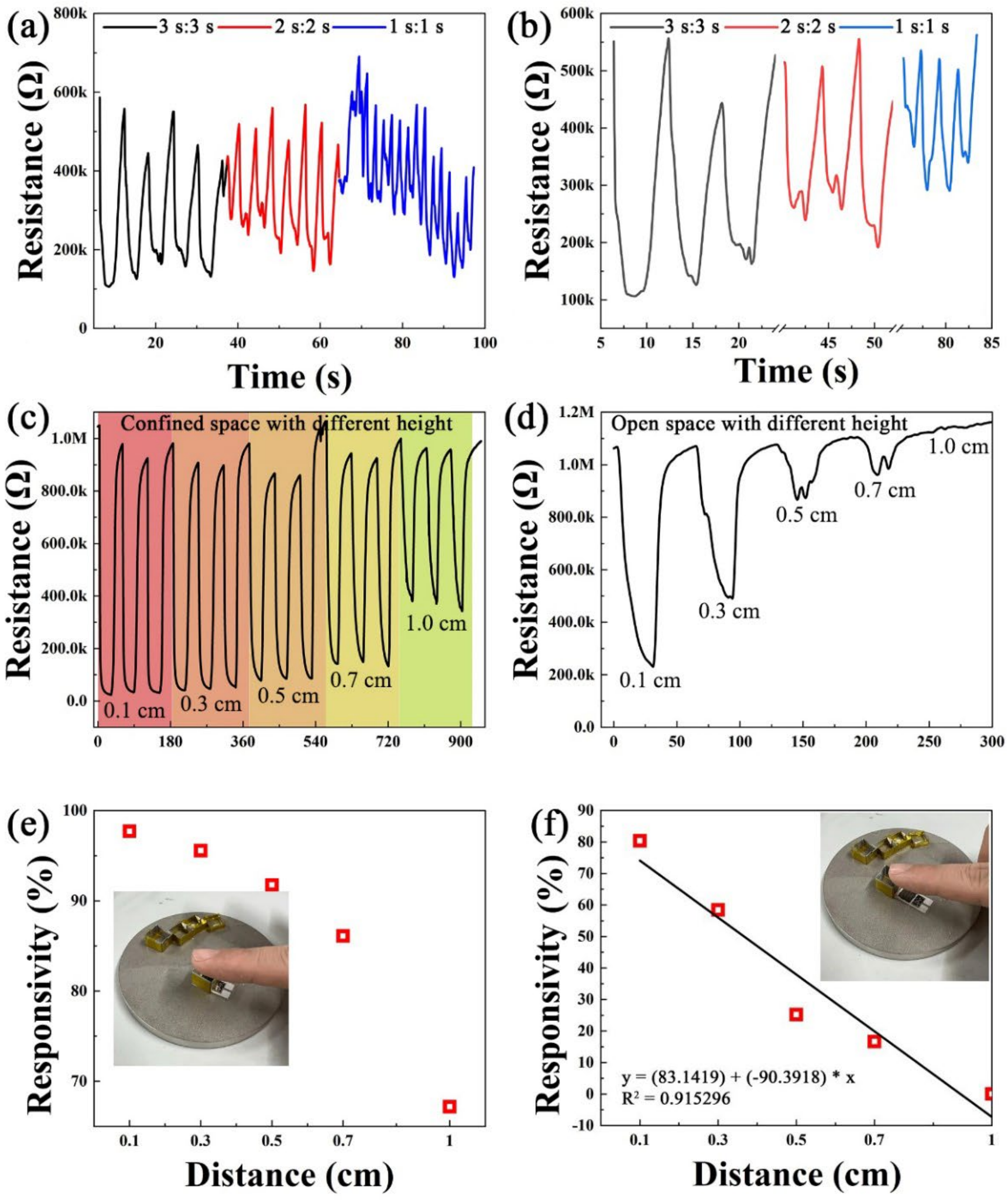

**Fig. 7** The prepared flexible humidity sensors based on 0.05 mL dispersion with $MoS_2$/GO volume ratio of 5:1 for respiration detection and fingertip proximity detection. (a) The curve of resistance of the prepared $MoS_2$/GO based humidity sensor changes with time during the processes of human respiration detection with different frequency of exhaling and inhaling. (b) Part of data point of (a) for clear display. (c, d) The curves of resistance of the prepared $MoS_2$/GO based humidity sensor changes with time during the processes of human fingertip proximity detection in a confined space

(c) and in an open space (d). (e, f) Responsivity versus distance during the processes of human fingertip proximity detection in a confined space (e) and in an open space (f).

**Conclusion**

In this study, the high-performance flexible humidity sensors based on $MoS_2$/GO composite films were fabricated in a simple and low-cost process, characterized and measured. We find that as the 0.05 mL dispersion with $MoS_2$/GO volume ratio of 5:1 was used for preparation of $MoS_2$/GO composite films, the humidity sensors show the best performance including the responsivity of about 98%, the response time of 1.3 s and recovery time of 12.1 s, excellent stability and repeatability. We applied the prepared flexible humidity sensors for the human respiration detection and fingertip proximity detection, showing excellent sensing performance. These findings will contribute to the rapid development of flexible humidity sensors based on graphene and related 2D materials.

**Author contributions**

Conceptualization, Gengwu Ge, Jie Ding, Wendong Zhang and Xuge Fan; investigation, Gengwu Ge; methodology, Gengwu Ge, Hongliang Ma and Xuge Fan; sample fabrication, Gengwu Ge and Ningfeng Ke; Characterization and measurements, Gengwu Ge; writing – original draft, Gengwu Ge and Xuge Fan; writing – review & editing, Hongliang Ma, Jie Ding, Wendong Zhang and Xuge Fan; supervision, Xuge Fan; project administration, Jie Ding and Xuge Fan; funding acquisition, Jie Ding and Xuge Fan.

**Corresponding Authors**

*E-mail: xgfan@bit.edu.cn, jie.ding@bit.edu.cn, wdzhang@nuc.edu.cn

**Conflicts of interest**

Xuge Fan, Jie Ding, and Gengwu Ge are co-inventors on a patent application (application number: 2024104436051) describing a method for highly sensitive and fast response flexible humidity sensors. The other authors have no competing financial interest.

**Acknowledgements**

This work was financially supported by the Beijing Natural Science Foundation (4232076), National Natural Science Foundation of China (62171037), 173 Technical Field Fund (2023-JCJQ-JJ-0971), National Key Research and Development Program of China (2022YFB3204600), National Science Fund for Excellent Young Scholars (Overseas), Beijing Institute of Technology Teli Young Fellow Program (2021TLQT012), Beijing Institute of Technology Science and Technology Innovation Plan. The SEM characterization was performed at the Analysis & Testing Center of Beijing Institute of Technology.